# Externally seeded backward lasing radiation from femtosecond laser filament in nitrogen gas

Pengji Ding[1], Sergey Mitryukovskiy[1], Aurélien Houard[1], Arnaud Couairon[2], André Mysyrowicz[1], and Yi Liu[1],*

[1] *Laboratoire d'Optique Appliquée, ENSTA ParisTech/CNRS/Ecole Polytechnique, 828, Boulevard des Maréchaux, Palaiseau, F-91762, France*
[2] *Centre de Physique Théorique, Ecole Polytechnique, CNRS, Palaiseau, F-91128, France*
*\*yi.liu@ensta-paristech.fr*

**Abstract**: Recently, S. Mitryukovskiy *et al*. presented experimental evidence showing that backward stimulated radiation at 337 nm can be obtained from plasma filaments in nitrogen gas pumped by circularly polarized 800 nm femtosecond pulses (Opt. Express, **22**, 12750 (2014)). Here, we report that this backward stimulated radiation is enhanced by a factor of ~ 16 in the presence of a seed pulse. This enhanced stimulated radiation can be either linearly or circularly polarized, dictated by the seeding pulse, which is distinct from the non-polarized nature of the ASE without seeding pulse. We also measured the spatial profile and estimated the energy of the radiation. This seeding effect confirms unambiguously the existence of population inversion between the $C^3\Pi_u$ and $B^3\Pi_g$ state of nitrogen molecules inside plasma filament and provides a possible solution to control the properties of this backward stimulated radiation.

## 1. Introduction

Stimulated radiation of air plasma pumped by ultrashort intense laser pulses has attracted growing attention in recent years [1-12]. Both backward and forward stimulated emission has been observed in experiments. In particular, the backward stimulated emission is very interesting, because it can be potentially employed for remote sensing applications. The employment of a backward stimulated lasing radiation for remote sensing is expected to bring tremendous improvement of measurement precision and sensitivity, because coherent detection methods such as SRS can be then used instead of the incoherent detection of laser induced luminescence [13].

Up to now, two different schemes of backward lasing action have been demonstrated. In the first method, a picosecond ultraviolet (UV) pulse (266 nm) was used to excite oxygen molecules in ambient air [1]. Population inversion between the $3p^3P$ and the $3s^3S$ states of oxygen atom was achieved by two photon dissociation of the oxygen molecules followed by resonant excitation of the atomic oxygen fragments. Both backward and forward Amplified Spontaneous Emission (ASE) at 845 nm has been observed in the experiments. However, application of this scheme for backward lasing generation for remote sensing is difficult due to the poor transmission of the UV pump pulse in atmosphere. Another scheme is based on population inversion in neutral nitrogen molecules. Backward stimulated emission from neutral nitrogen molecules inside a laser plasma filament was first suggested in 2003, based on the observed exponential increase of the backward UV emission with the filament length [2]. In 2012, D. Kartashov and coworkers focused a mid-infrared femtosecond laser pulse (3.9 μm or 1.03 μm) inside a high pressure mixture of argon and nitrogen gas. Backward stimulated emissions at 337 nm and 357 nm were observed with an optimal argon gas pressure of 5 bar and nitrogen pressure of 2 bar [3]. The emission at 337 nm and 357 nm have been identified as being due to the transition between the third and second excited triplet states of neutral nitrogen molecules, *i.e.* $C^3\Pi_u \rightarrow B^3\Pi_g$. The population inversion mechanism between the $C^3\Pi_u$ and $B^3\Pi_g$ states has been attributed to the traditional Bennet mechanism, where collisions transfer the excitation energy of argon atoms to molecular nitrogen [14]. Unfortunately, this method cannot be applied for remote generation of backward lasing emission because of its requirement of high pressure argon gas ($p > 3$ bar).

A few months ago, S. Mitrykovskiy *et al.* showed that a backward ASE at 337 nm can be obtained from laser filaments pumped by circularly polarized 800 nm femtosecond laser pulses [6]. Based on the critical dependence of the 337 nm radiation on pump laser polarization and the distinct dependence of the 337 nm radiation on pump laser energy for circularly and linearly polarized pump pulses, population inversion between the $C^3\Pi_u$ and $B^3\Pi_g$ states was obtained in pure nitrogen. The authors attributed the mechanism of population inversion to the inelastic collision between the energetic electrons and the neutral nitrogen molecules [6]. However, due to the weakness of the backward ASE, the beam properties of this backward radiation such as its divergence and energy have not been characterized.

In this paper, we report that an external seeding pulse around 337 nm in the backward direction leads to a ~ 16 time enhancement of the backward ASE. At the same time, the divergence of the seeded lasing is found to be significantly reduced compared to that of ASE. Moreover, the seeded lasing radiation inherits the polarization property of the seed pulse. These three observations confirm unambiguously the previous assumption of population inversion between the relevant nitrogen molecular states. The critical role of pump laser ellipticity was also observed in this seeded backward lasing scheme, which supports the hypothesis that inelastic collision between the energetic electrons with neutral nitrogen molecules is at the origin of population inversion.

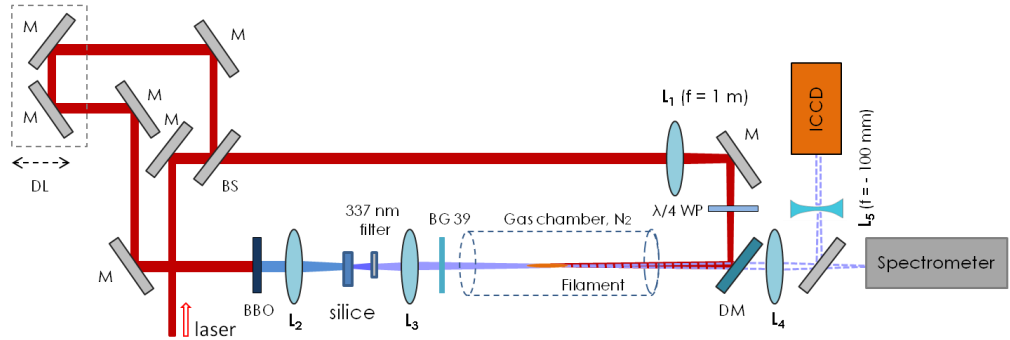

Fig. 1. Schematic experimental setup.

## 2. Experimental setup

In our experiment, a commercial Chirped Pulses Amplification (CPA) laser system (Thales Laser, Alpha 100) was used. This system delivers 40 fs laser pulses at a repetition rate of 100 Hz, with maximum pulse energy of 12 mJ. A schematic experimental setup is presented in Fig. 1. The output laser pulse was split into a main pump pulse and a much weaker second pulse by a 1 mm thick 5%/95% beamsplitter. The pump pulse passed through a λ/4 waveplate and then was focused by an $f = 1000$ mm convex lens ($L_1$). A dichromatic mirror was used to reflect the focused 800 nm pump pulses into a gas chamber filled with pure nitrogen gas at 1 bar pressure. This dichromatic mirror reflects more than 99% of the 800 nm pump pulse and it is transparent to the backward ultraviolet emission from the laser plasma situated inside the gas chamber. The second weaker 800nm pulse first passed through a mechanical delay line and then through a 1 mm thick type-I BBO crystal in order to generate femtosecond pulses at 400 nm. The 400 nm pulse was linearly polarized in the vertical direction. The obtained 400 nm pulse was further focused by an $f = 200$ mm convex lens ($L_2$) inside a 20 mm long fused silica sample to broaden its spectrum through intense nonlinear interaction. We selected the spectrum component around 337 nm with an interference bandpass filter, which has a transmission peak at 340 nm and a bandwidth of 10 nm. The resulting pulse centered at 340 nm, referred to as seeding pulse in the following, was focused by another $f = 100$ mm lens into the gas chamber from the opposite direction of the pump pulses. The separation between the lenses $L_2$ and $L_3$ was adjusted to insure that the geometrical focus of the seeding pulse overlapped with the central part the of the long plasma filament in the longitudinal direction. The transverse spatial overlapping between the geometrical focus of the seeding pulse and the center of the plasma filament was carefully assured by translating finely the focal lens ($L_3$) in the transverse plane. The temporal delay between the 800 nm pump pulses and the seeding pulse at 337 nm could be adjusted by the mechanical delay line. For some of our experiments, we installed a $\lambda/4$ waveplate for 400 nm radiation after the BBO crystal so that a circularly polarized seeding pulse could be obtained after filamentation inside the fused silica sample. The backward emission from the laser plasma filaments was detected by either a spectrometer (Ocean Optics HR 4000), an intensified Charge Coupled Device (iCCD) camera (Princeton Instrument, model: PI-MAX), a calibrated photodiode, or a sensitive laser power meter (model: OPHIR, NOVA, PE9-C).

### 3. Experimental results and discussion

We first measured the spectra of the backward emission from the laser plasma with/without the seeding pulses, for both circularly and linearly polarized pump lasers. The results are presented in Fig. 2. In the case of circularly polarized pump pulses, a sharp radiation peak at 337 nm can be observed even without external seed pulse, which is referred to as a backward ASE in our previous work [6]. With the external seed pulse, an enhancement of the peak by a factor of ~ 16 is found (Fig. 2 (a)). Considering the intensity of the seed pulse at the 337 nm spectral position, we estimated that the seed pulse is amplified by a factor of $2.5 \times 10^5/2000 = 125$ times. In the case of linearly polarized pump pulses (Fig. 2 (b)), no detectable ASE was observed and no amplification of the seed pulse could be observed. These observations confirm our previous conclusion that population inversion responsible for the stimulated 337 nm radiation is only established with circularly polarized pump pulses [6].

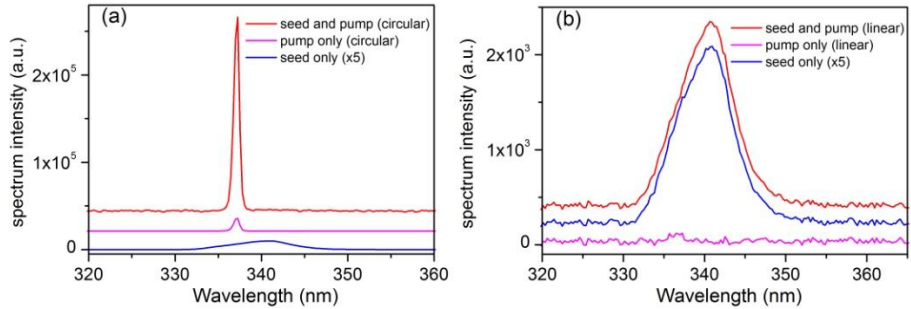

Fig. 2. Spectra of the backward stimulated emission with circularly (a) and linearly (b) polarized pump pulsed at 800 nm. The spectra of the seed pulses and those of the backward emission from just the pump pulses are also presented for comparison.

We then studied the polarization property of this seeding effect by injecting linearly and circularly polarized seeding pulses inside the plasma filament. In order to analysis the polarization properties of the lasing radiation, we installed a Glan-Taylor prism before the detecting photodiode. In the experiment, we recorded the intensity of the transmitted 337 nm radiation as a function of the rotation angle of the Glan-Taylor prism. The result for the ASE obtained without seeding pulse is first presented in Fig. 3 (a), indicating that the ASE is not polarized. For linearly polarized seed pulses in the vertical direction, we observed that the amplified lasing signal is also linearly polarized in the same direction (see Fig. 3 (b)), evidenced by the good agreement between the experimental results and the theoretical fit with Malus' law. The result for circularly polarized seeding pulses is presented in Fig. 3 (c), where a circularly polarized amplified emission is also observed. The maintenance of the pulse polarization during the amplification is in agreement with our hypothesis that that population inversion is present and responsible for the seeding pulse amplification.

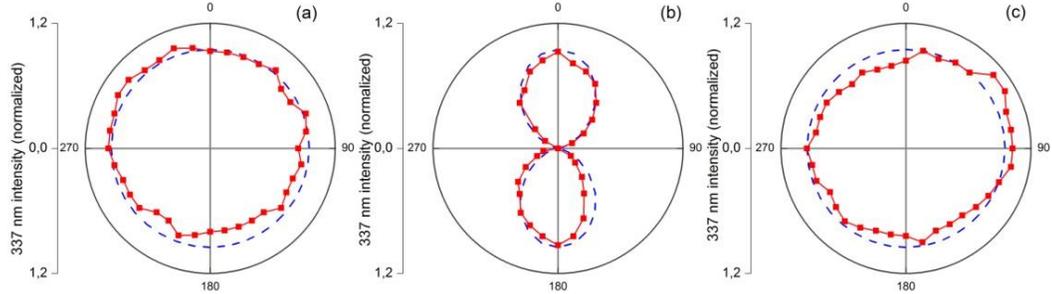

Fig. 3. Polarization properties of the ASE (a), the seeded backward stimulated emission with linearly (b) and circularly (c) polarized seed pulses at 337 nm. The dots present the experimental results and the dashed lines denote the theoretical fitting.

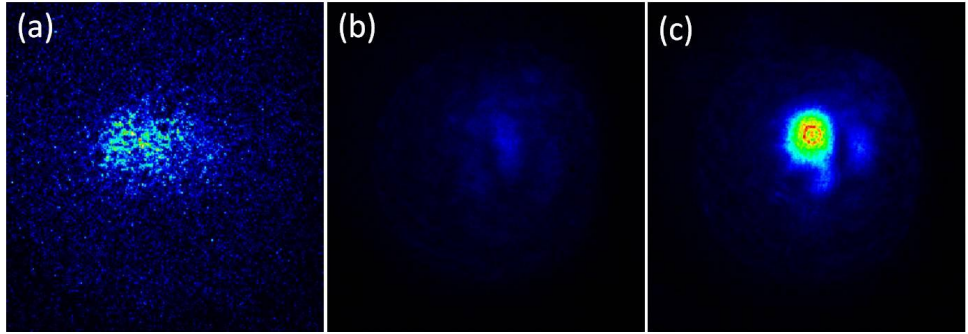

Fig. 4. Spatial profile of the backward ASE (a), the seed pulse (b), and the seeded stimulated radiation (c). The opening angle of each panel is 24 mrad × 24 mrad.

We also measured the spatial profile of the ASE and the seeded lasing emission with the iCCD camera. In Fig. 4 (a), the measured spatial profiles of the backward emission without seeding pulse are shown for circularly pump pulses of 800 nm. The ASE exhibits a Gaussian distribution with a divergence of 9.2 mrad (Fig. 4 (a)). In the case of linearly polarized 800 nm pump pulse, no backward emission at 337 nm was observed with the iCCD. These spatial profile measurements agree with the above spectral measurements where significant stimulated emission at 337 nm can only be observed with circularly polarized pump pulses. We present the spatial profile of the seeding pulse in Fig. 4 (b). This weak pulse exhibits a divergence around 20 mrad in our experiments. In the presence of both pump and seeding pulse, an extremely intense 337 nm radiation was found, as presented in Fig. 4 (c). This amplified stimulated emission shows a divergence angle of ~ 3.8 mrad, much smaller than that of the ASE and the seeding pulse.

We have also measured the pulse energy of the seeded lasing radiation with a sensitive laser energy meter. The energy of seeded lasing pulse was measured to be around 5.0 nJ, corresponding to an energy conversion efficiency of $5.1 \times 10^{-7}$ from 9.8 mJ pump pulses. Therefore, the pulse energy of the ASE was deduced to be 310 pJ with a conversion efficiency of $3.2 \times 10^{-8}$.

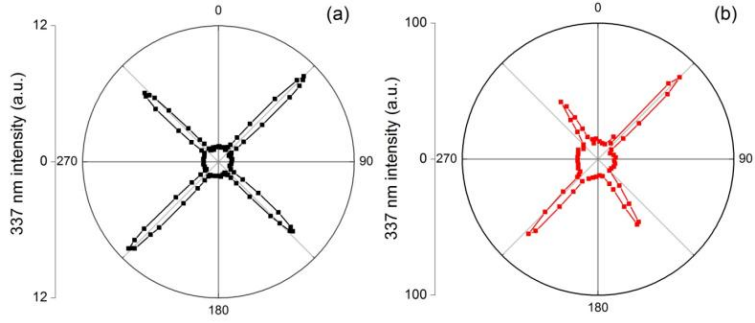

Fig. 5. Dependence of the backward ASE (a) and the seeded backward stimulated radiation (b) as a function of the rotation angle of the quarter-wave plate. The angles $\varphi = 90° \times m$ correspond to linearly polarized laser, with $m = 0, 1, 2, 3$. The angles $\varphi = 45° + 90° \times m$ correspond to circularly polarized laser.

All the above experimental observations highlight the crucial role of pump laser polarization. To evaluate that, we measured systematically the seeded lasing emission intensity by rotating the λ/4 waveplate for the pump pulses. In Fig. 5 (a), the result for the ASE without seeding pulse is first presented as a function of the rotation angle $\phi$ of the waveplate, which has been reported in our previous work [6]. Intense ASE was observed only with circularly polarized pump pulses and shows dramatic decrease when the ellipticity deviates from $\varepsilon = 1$. In the presence of a constant linearly polarized seeding pulse, a similar dependence on laser ellipticity was observed (Fig. 5 (b)). This confirms that population inversion between the $C^3\Pi_u$ and $B^3\Pi_g$ states can be only achieved with circularly polarized pump pulses. The slight asymmetry and the deviation of the peaks from $\phi = 135°$ and $\phi = 315°$ can be due to the fact that the circularly polarized pump pulses reflect on the dielectric dichromatic mirror in this experiment. This mirror has slightly different reflectivity for $p$- and $s$- polarized light and thus results in a non-perfect circularly polarized pump pulses after reflection.

In our previous work, we have attributed the population inversion to the following inelastic collision process:

$$N_2(X^1\Sigma_g^+) + e = N_2(C^3\Pi_u^+) + e.$$

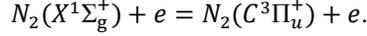

The effectiveness of circular laser polarization for population inversion lies in the fact that photoelectrons generated in a circularly polarized laser field are left with a substantial kinetic energy after the passage of the pump laser pulse. In our case of laser intensity around $1.4 \times 10^{14}$ W/cm$^2$ [6], most of the electrons obtain kinetic energy around 16 eV, which is sufficiently high to excite the ground state nitrogen molecules to the third triplet state through inelastic collisions [15].

Finally, we point out that all the above experiments were performed in pure nitrogen. We have previously found that the presence of oxygen molecules deteriorates the backward ASE [6]. In this seeded lasing scheme, we observed no significant backward lasing signal in ambient air. This indicates that the detrimental role of oxygen molecules is due to its ability to ruin population inversion, which was attributed to the collision quenching process [6].

## 4. Conclusion

In conclusion, we demonstrated that the backward ASE from plasma filaments, which is generated in nitrogen gas pumped by circularly polarized 800 nm femtosecond laser pulses, can be enhanced by a factor of ~ 16 in the presence of a backward seeding pulse. The amplified lasing radiation inherits the polarization property of the seeding pulse and its divergence angle was found to be around 3.8 mrad, much less than that of the backward ASE. The critical role of pump laser polarization was also observed in the seeded lasing regime, where intense lasing effect was only possible for circularly polarized pump pulses. The amplification phenomenon, the reduced divergence of the seeded lasing radiation, and the critical role of pump laser polarization confirms unambiguously the presence of population inversion between the $C^3\Pi_u$ and $B^3\Pi_g$ states of neutral $N_2$ molecules in the filament plasma suggested previously. At the same time, this external seeding scheme provides a possible method to boost the energy of the backward lasing radiation, which is important for future applications.